%% file: Global.tex
\documentclass[floatfix,aps,prc,noeprint,twocolumn]{revtex4-1} % two column journal format

\usepackage{amsfonts}                            % Use AMS fonts
\usepackage{amssymb}                             % and symbols
\usepackage{amsmath}  
\usepackage{mathtools}                           % and definitions
\usepackage{bm}
\usepackage{graphicx}                            % Graphics package
\usepackage{color}
\usepackage{latexsym}                            % Load additional symbols
\usepackage{dcolumn}                             % Decimal-dot aligned columns
\usepackage[colorlinks=true,linkcolor=blue,citecolor=blue,urlcolor=blue]{hyperref}
\usepackage[normalem]{ulem}
\usepackage{slashed}
\usepackage{xfrac}
\usepackage{cleveref}
\crefname{figure}{Fig.}{Figs.}
\crefname{equation}{Eq.}{Eqs.}
\usepackage{natbib}
\usepackage{mathrsfs}
\usepackage{physics} 
\usepackage{comment}
\usepackage[caption=false]{subfig}
\newcommand\be{\begin{equation}}
\newcommand\ee{\end{equation}}
\newcommand\bea{\begin{eqnarray}}
\newcommand\eea{\end{eqnarray}}
\newcommand\bes{\begin{subequations}}
\newcommand\ees{\end{subequations}}
\begin{document}

\title{A novel approach to the global analysis of proton form factors in elastic electron-proton scattering}

\author{C.~M$^\textrm{c}$Rae and P.~G.~Blunden}
\affiliation{\mbox{Department of Physics and Astronomy,
	University of Manitoba}, Winnipeg, Manitoba, Canada R3T 2N2}

\begin{abstract}
We present a novel method for the global analysis of elastic electron-proton scattering when combining data sets from experiments with different overall normalization uncertainties.
The method is a modification of one employed by the NNPDF collaboration in the fitting of parton distribution data.
This method is an alternative to the `penalty trick' method traditionally employed in global fits to proton electric and magnetic form factors, while avoiding the biases inherent in that approach.
We discuss issues that arise when extending the method to nonlinear models.
For data with $Q^2>1~\textrm{GeV}^2$ we find relatively minor differences to traditional model fits when the normalization uncertainties from different experiments are correctly accounted for.
We discuss implications of this method for the well-known discrepancy between the form factor ratio $G_E/G_M$ extracted from the Rosenbluth and polarization transfer techniques.
\end{abstract}

%\date{\today}
\maketitle
%% ----------------------------------------------------------------
\section{Introduction}
Elastic electron-proton scattering can be described by two functions of the squared four-momentum transfer $Q^2$: the Sachs electric and magnetic form factors $G_E\left(Q^2\right)$ and $G_M\left(Q^2\right)$, respectively.
Experimentally there are two main techniques to determine the form factors: a longitudinal-transverse (LT) separation of unpolarized cross section data~\cite{Rosenbluth:1950}, and polarization transfer (PT) measurements of the `polarization ratio' $G_E/G_M$.
After taking into account standard radiative corrections, separate fits of the form factors to these data lead to fits which disagree significantly with one another as $Q^2$ increases.
Model-dependent two-photon exchange (TPE) corrections to LT data are considered the most viable explanation for this discrepancy~\cite{Blunden:2003, Guichon:2003}, but definitive experimental evidence at high $Q^2$ is lacking.

Under the one-photon exchange (OPE) assumption the differential cross section for elastic electron-proton scattering is proportional to the reduced cross section
\be
\label{eq:reduced}
\sigma_\textrm{red} = \varepsilon\, G_E^2\left(Q^2\right)+\tau \,G_M^2\left(Q^2\right),
\ee
where $\tau = Q^2/(4 M^2)$ and $\varepsilon^{-1} = 1+2(1+\tau) \tan^2{\theta/2}$.
Here $\theta$ is the scattering angle in the lab frame, $M$ is the proton mass, and the electron is taken to be massless.

The traditional Rosenbluth LT separation~\cite{Rosenbluth:1950} of $G_E$ and $G_M$ uses measurements of $\sigma_\textrm{red}$ at fixed $Q^2$ with a linear fit to the virtual photon polarization $\varepsilon$ by varying the scattering angle.
This produces form factors that are model-independent but are of limited range in beam energy and scattering angle.
Alternatively, one can make a global fit of models of $G_E$ and $G_M$ to cross section data over a wide range of kinematical conditions~\cite{Bernauer:2014}.
This second method often requires combining data sets from different experiments, or different experimental setups within the same experiment, with different absolute normalizations and associated uncertainties.

A key question, which we address in this paper, is how to combine such measurements from different data sets in a statistically rigorous and meaningful way.
To do so we introduce a novel method based on a modification of one employed by the NNPDF collaboration~\cite{Ball:2010}, which we call the iterated model fit (IMF).
As a proof of concept, we compare this method to two recent analyses of cross section data that use the traditional `penalty trick' method, which we will describe below.
The first analysis is that of Gramolin and Nikolenko~\cite{Gramolin:2016}, who use a linear parameterization for $\sigma_\textrm{red}$ in the model functions $G_E^2$ and $G_M^2$.
They use three experimental data sets~\cite{Walker:1994, Andivahis:1994} with $Q^2$ in the range from 1 to 8.83~GeV$^2$.
The second analysis is by the GMp12 collaboration~\cite{Christy:2022}, who use the same three data sets plus six additional data sets, including their own new precision data with $Q^2$ up to 15.75~GeV$^2$.
They use a more flexible parameterization of model functions for $G_M$ and $(\mu_p G_E/G_M)^2$, which leads to a nonlinear model for $\sigma_\textrm{red}$.
We discuss how the IMF method can be adapted to such a nonlinear fitting procedure.

References~\cite{Gramolin:2016, Christy:2022} both apply updated and improved radiative corrections to the original experimental data, but do not include TPE corrections.
Both groups still find significant disagreement between separate fits of the cross section and PT data.
As it is the intention of this paper to compare methodologies rather than construct a state-of-the-art global fit to $G_E$ and $G_M$, we leave it to future work to incorporate TPE contributions as well as cross section data with $Q^2 < 1~\textrm{GeV}^2$.

\section{Fitting with normalization uncertainty}
The technique of fitting via minimization of a chi-square emerges out of the more fundamental desire to find the parameters of a model which, assuming the model is correct, are most likely to have produced the observed data (a maximum likelihood estimate).
Thus if we are not careful when we modify the chi-square to be minimized, we will violate the underlying assumption that measurements are normally distributed (the chi-square is by definition the exponent of a Gaussian likelihood), leading to undesirable biases in our fitted models. 

The standard method for dealing with normalization uncertainty is fitting via a modification of the traditional chi-square known colloquially as the `penalty trick',
\begin{equation}
\label{eq:chi_penalty_updated}
\begin{split}
\chi^2_\mathrm{p}
 &= \sum_{i=1}^{\mathcal{N}} \left[\frac{\left(n_{i}-1\right)^2}{\left(\Delta n_{i}\right)^2}+\sum_{j=1}^{N_{i}} \frac{\left(y_{i j}-\operatorname{M}_{ij}/n_i\right)^2}{\left(\Delta y_{i j}\right)^2}\right].
\end{split}
\end{equation}
Here a model $\operatorname{M}$ (in our case $\operatorname{M}=\sigma_\textrm{red}$) is fitted, alongside floating normalizations ($n_i$), to data ($y_{ij}$) with both additive ($\Delta y_{ij}$) and multiplicative ($\Delta n_i$) uncertainties; $i$ indexes the $\mathcal{N}$ experiments, and $j$ indexes the $N_i$ data points of each experiment (in more naive approaches, one may scale only the data by the $n_i$).
This method fails to be Gaussian in the normalization parameters $n_i$, so we seek a more sound alternative. 

In a seminal paper by Ball {\em et al.}~\cite{Ball:2010}, the NNPDF collaboration firstly demonstrate the biases in the usual approaches for combining multiple experimental data sets with multiplicative uncertainties, and secondly, for fitting a single parameter, construct a Monte Carlo method which takes the normalization uncertainty into account without compromising the integrity of the fit. 
The most important steps in the derivation of the method are as follows.
Firstly, perform the usual error propagation for multiplying data with uncertainty:
\begin{equation}
(y_{ij} \pm \Delta y_{ij})\,(1\pm \Delta n_i) = y_{ij} \pm \sqrt{( \Delta y_{ij})^2+(y_{ij} \Delta n_i)^2} .
\end{equation}
Secondly, recognizing that a well fitted model will resemble the data; in the chi-square of that error-propagated data, replace the instance of the data in the propagated error with a best guess for the model, $\hat{\operatorname{M}}$: 
\begin{equation}
\label{eq:finalsimplechi}
\chi^2_{t_0} = \sum_{i=1}^{\mathcal{N}}\left[\sum_{j=1}^{N_{i}} \frac{\left(y_{i j}-\operatorname{M}_{ij}\right)^2}{\left(\Delta y_{i j}\right)^2+\left(\hat{\operatorname{M}}_{i j}\Delta n_i\right)^2}\right].
\end{equation}
This $\chi^2_{t_0}$ has all the nice statistical properties we expect, and its likelihood remains Gaussian in both model and data.
If our best guess coincides with the model given from the optimization, the replacements are justified and we are done.
Else, we use the optimized model as our ensuing guess, iterating until the model converges.
This is the $t_0$ method of Ref.~\cite{Ball:2010}, which focuses on fitting single parameters at specific kinematics.

When one attempts to apply the $t_0$ method to data at multiple kinematics, it is ambiguous for off-diagonal elements of the constructed covariance matrix, which kinematics the best-guess model should be evaluated at.
We propose to generalize the method by appending the original data covariance matrix $\textrm{cov}_i$ as follows for each of the $i$ experiments:
\begin{equation}
\label{eq:covIMF}
\textrm{cov}^\textrm{IMF}_{i,jk} = \textrm{cov}_{i,jk} + (\Delta n_i)^2 \operatorname{\hat{M}}_{ij}\operatorname{\hat{M}}_{ik}.
\end{equation}
This generalizes the case of a single data point per experiment and of all data points being at the same kinematics~\cite{Ball:2010}.
With this, the remaining step used by NNPDF is left unaltered, and replica data are generated via a two step Monte-Carlo procedure.
Firstly, for each datum one generates replica data $y^R_{ij}$ by pulling from a normal distribution characterized by the point's central value and uncertainty.
Secondly, a normalization factor $n^R_{i}$ is pulled from a normal distribution centered around $1$ with a standard deviation defined by the normalization uncertainty for the datum's experiment.
Thus the final chi-square to minimize during our Monte Carlo replica iterated model fit is given in matrix form by
\begin{equation}
\label{eq:chiIMF}
    \!\!\chi^2_\textrm{IMF} = \sum_{i=1}^{\mathcal{N}}\!\left(n^R_i y^R_i -\operatorname{M}_i\right)^\intercal\!(\textrm{cov}^\textrm{IMF}_{i})^{-1}\!\left(n^R_i y^R_i- \operatorname{M}_i\right)\!.
\end{equation}
For $\mathcal{R}$ replica's generated we will have $\mathcal{R}$ best fit models and $\mathcal{R}$ sets of best fit parameters.
These fits are averaged to produce an updated best fit model $\operatorname{\hat{M}}$ of the original data, which is then fed back into Eq.~(\ref{eq:covIMF}).
This process iterates until we find convergence of the output model.

\subsection{IMF vs.~penalty trick (linear model)}
Here we compare fits via the penalty trick method to fits via the IMF method to understand how the ubiquitous method compares to the statistically rigorous method.
Gramolin and Nikolenko~\cite{Gramolin:2016} use a simple linear model for the square of the form factors to ensure the fitting procedure becomes a linear algebra problem in the parameters (intentionally utilizing the more naive version of the penalty trick). 
They take
\bea
\hspace{-2em}
G_{E}^2(Q^2; \bm{a}) &=&\left(1-a_{1} \tau-a_{2} \tau^2-a_{3} \tau^3\right) G_\mathrm{dip}^2,\nonumber\\
G_{M}^2(Q^2; \bm{b})/\mu_p^2 &=&\left(1-b_{1} \tau-b_{2} \tau^2-b_{3} \tau^3\right) G_\mathrm{dip}^2,
\label{eq.Model_Gram}
\eea
where $\mu_p$ is the proton magnetic moment, and $G_\mathrm{dip}(Q^2)$ is a dipole form factor.
The downside to using such a simple model is manyfold.
The squared form factors have nothing preventing them from becoming negative, the form factors do not fall off as predicted by QCD scaling laws, and if one wishes to include PT data the fitting procedure becomes nonlinear.
Nevertheless, it makes for a rather simple point of comparison.
For the penalty trick method we solve the system of equations $\partial \chi^2_{\mathrm{p}}/\partial \alpha_i=0$ for the nine parameters $\alpha_i$, giving the results shown in Table~\ref{tab:GramFit}.
The index of $n_i$ refers to the Walker and Andivahis (8 GeV and 1.6 GeV) data sets itemized in Ref.~\cite{Christy:2022}.

\begin{figure}[t]
\centering
    \includegraphics[width=\columnwidth]{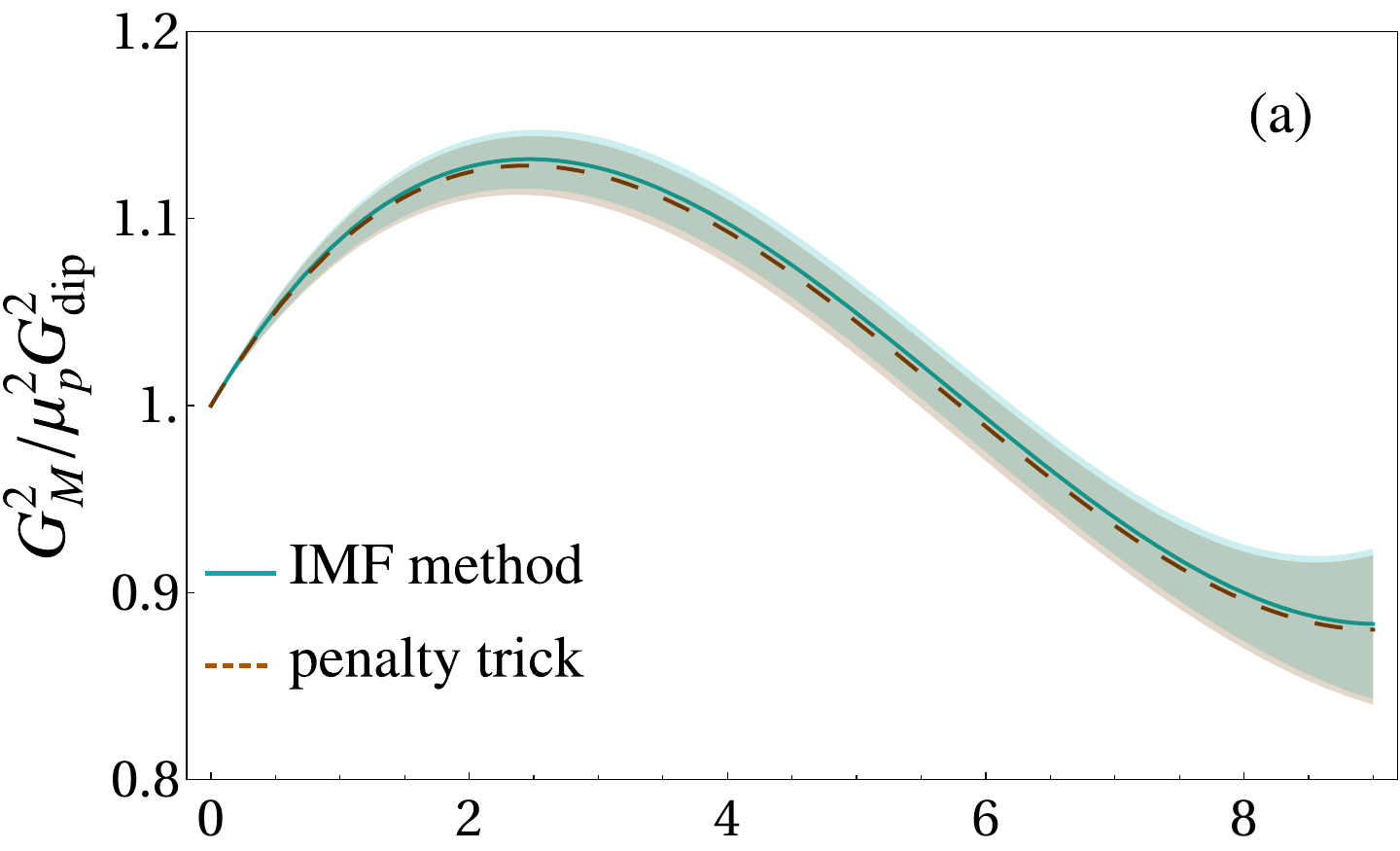}
    \includegraphics[width=\columnwidth]{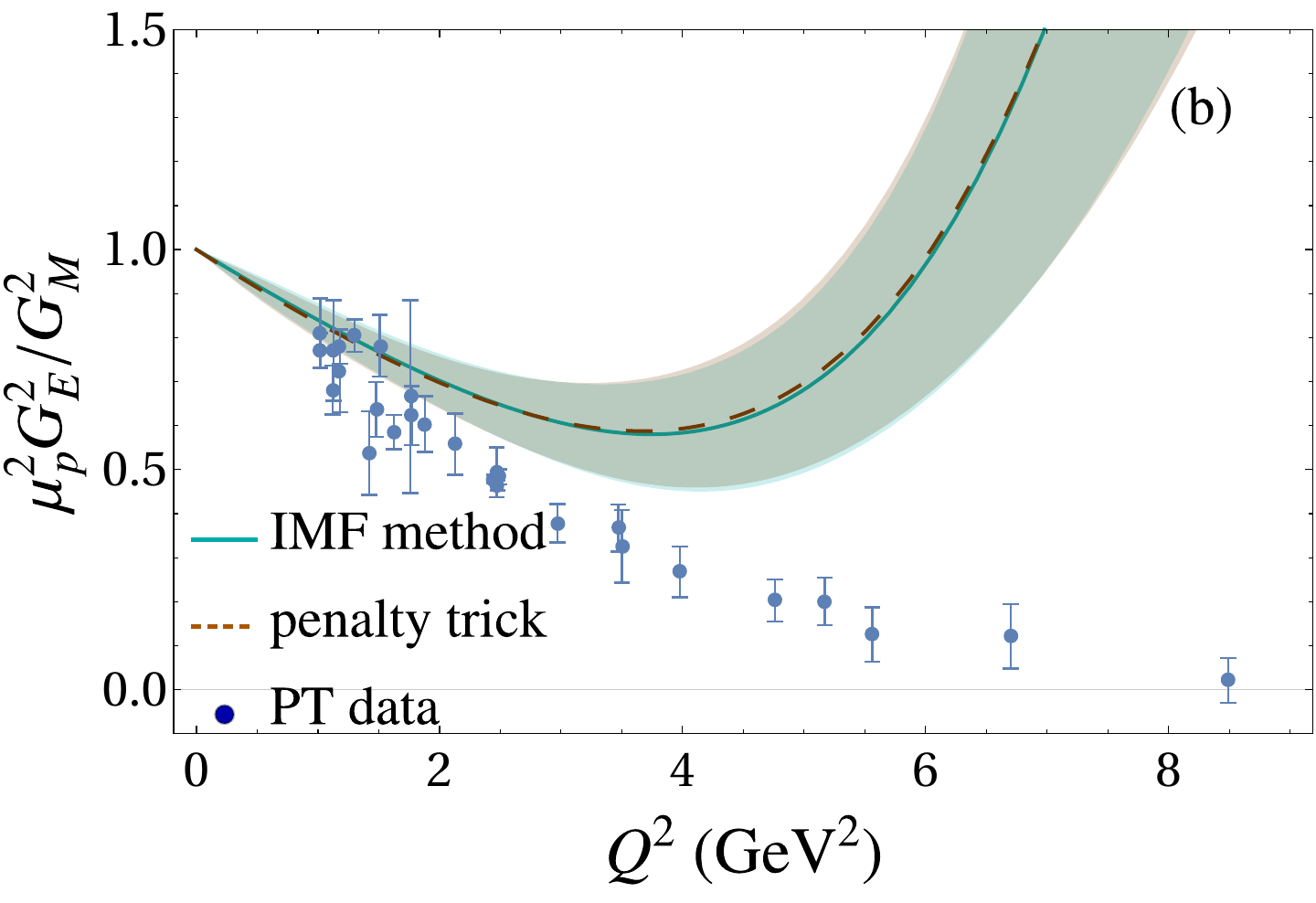}
%\vspace{-3ex}
\caption{(a) The normalized $G_M^2(Q^2)$ for both the IMF method (solid line) and the penalty trick (dashed line), and (b) the squared form factor ratio $\mu_p^2 G_E^2/G_M^2$, plotted with PT data itemized in the Supplementary Material of Ref.~\cite{Christy:2022} with $Q^2 > 1~\textrm{GeV}^2$ (not included in the fit).
}
\label{fig:FFGram}
\end{figure}

To implement the IMF method we begin with a guess of all parameters being $0$, which sets the form factors to standard dipoles.
Convergence required only 2 iterations of the IMF procedure, with $\mathcal{R} = 1000$ replica data sets fitted at each iteration.
The parameters in the column labeled IMF in Table~\ref{tab:GramFit} are the mean and standard deviation of the final six dimensional distribution of 1000 fitted parameters.
In Fig.~\ref{fig:FFGram} we can see plots of the  $G_M^2$ and $G_E^2/G_M^2$ form factor fits, suitably normalized.

\begin{table}[t]
\renewcommand\arraystretch{1}
\centering
\caption{Form factor parameters extracted from the penalty trick and iterated model fit (IMF) methods for the linear model of Eq.~(\ref{eq.Model_Gram}).
We use the same three data sets as Ref.~\cite{Gramolin:2016}, but with updated radiative corrections from Ref.~\cite{Christy:2022}.}
\begin{tabular*}{\columnwidth}{l @{\extracolsep{\fill}} cc}
\hline\hline
& \multicolumn{1}{c}{penalty trick} & \multicolumn{1}{c}{IMF}\\
 \hline
 $a_1$ & $~~0.23\pm 0.22$ & $~~0.17 \pm 0.22$\\
 $a_2$ & $~~0.47\pm 0.44$ & $~~0.55\pm 0.44$\\
 $a_3$ & $-0.35\pm 0.21$ & $-0.38\pm 0.21$\\
 \noalign{\vskip 1ex}
 $b_1$ & $-0.407\pm 0.045$ & $-0.413\pm 0.045$\\
 $b_2$ & $~~0.373\pm 0.045$ & $~~0.375\pm 0.046$\\
 $b_3$ & $-0.076\pm 0.013$ & $-0.077\pm 0.013$\\
\hline
 $n_1$ & $1.008\pm 0.012$ & ---\\
 $n_2$ & $1.008\pm 0.012$ & ---\\
 $n_3$ & $0.963\pm 0.013$ & ---\\
\hline
 $\chi^2/\mathrm{d.o.f}$ & $28.3/39$ & $28.1/42$\\
\hline\hline
\end{tabular*}
\label{tab:GramFit}
%\vspace{-3ex}
\end{table}

\subsection{IMF vs.~penalty trick (nonlinear model)}
The application of the IMF method to nonlinear models creates a potential difficulty.
Generically, the average of several functions which share the same functional form $F(x;\bm{\alpha})$ will only maintain that functional form when the function is linear in $\bm{\alpha}$.
This ruins the iterative step because we do not have the same functional form at each iteration.
There are two possible ways out of this.
Firstly, one could find the parameters $\bm{\alpha}$ for the chosen functional form which are closest to the average of the replica fits $\langle F_R(x)\rangle$ by minimizing the $L^2$ norm
\be
\frac{\partial}{\partial \bm{\alpha}} \int_{x_\textrm{min}}^{x_\textrm{max}} dx\ 
\Bigl(F(x,\bm{\alpha}) - \langle F_R(x)\rangle \Bigr)^2 = \bm{0}.
\ee
This minimization is computationally expensive, but it ensures that the functions are similar as possible over the relevant range.
Alternatively, one may simply keep taking average parameters and plugging them into the model, understanding that this is only an approximation for the aforementioned closest model to the average model, and hope for convergence, as convergence is all one needs for the results to be statistically sound. 
The second option works for the following model, and is used throughout the remainder of this paper. 

A flexible parametrization for the form factors is the $z$-expansion model presented in Ref.~\cite{Lee:2015}.
The model is
\bea
G_E\left(z;\bm{a}\right) &=& \sum_{k=0}^{k_\textrm{max}} a_{k} z^{k}, \quad
G_M\left(z; \bm{b}\right)/\mu_p = \sum_{k=0}^{k_\textrm{max}} b_{k} z^{k},\nonumber
\\
z(Q^2) &=& \frac{\sqrt{t_\mathrm{cut}+Q^2}-\sqrt{t_\mathrm{cut}-t_0}}
{\sqrt{t_\mathrm{cut}+Q^2}+\sqrt{t_\mathrm{cut}-t_0}}.
\label{eq:zexpmod}
\eea
In all fits we use the values $t_{\mathrm{cut}}=4 m_\pi^2$ and $t_0=t_\mathrm{cut}\left(1-\sqrt{1+Q_\textrm{max}^2/t_\textrm{cut}}\right)$, but the fits are rather insensitive to these particular values.
To ensure that $G_E(Q^2)\sim 1/Q^4$ as $Q^2\to\infty$ there are constraints
\be
\sum_{k=n}^{k_\textrm{max}} \frac{k!}{(k-n)!} a_{k}=0, \quad n=0,1,2,3 ,
\ee
and similarly for $b_k$.
Additionally we require $G_E(0) = 1$ and $G_M(0)=\mu_p$.
Choosing $k_\textrm{max} = 8$ therefore leaves 4 free parameters for each form factor, which we designate to be those with $k=0,1,2,3$.

For the following fits we use the global cross section data from the GMp12 analysis~\cite{Christy:2022} with $Q^2<16~\mathrm{GeV}^2$ (115 points, only six points are excluded).
This compilation was built specifically from experiments for which past radiative corrections could be undone and then recalculated with updated corrections to ensure the analysis was consistent across all the data.
The GMp12 analysis uses the penalty trick to account for the normalization uncertainty in their fits, but with a different parameterization than given in Eq.~(\ref{eq:zexpmod}).
Implementing the IMF method, and comparing against PT data, gives us another clean comparison between the fitting procedures. 

\begin{table}[t]
\renewcommand\arraystretch{1}
\centering
\caption{Form factor parameters from the penalty trick and IMF methods for the nonlinear model of Eq.~(\ref{eq:zexpmod}).
Fitted normalizations for the nine data sets (not shown) range from $0.95$ to $1.06$, consistent with Ref.~\cite{Christy:2022}.
}
\begin{tabular*}{\columnwidth}{l @{\extracolsep{\fill}} cc}
\hline\hline
& \multicolumn{1}{c}{penalty trick} & \multicolumn{1}{c}{IMF}\\
 \hline
 $a_0$ & $-0.847 \pm 0.028$ & $-0.847\pm 0.028$\\
 $a_1$ & $~~1.269 \pm 0.065$ & $~~1.269\pm 0.067$\\
 $a_2$ & $~~0.47 \pm 0.34$ & $~~0.45\pm 0.34$\\
 $a_3$ & $-2.36 \pm 0.49$ & $-2.31\pm 0.51$ \\
 \noalign{\vskip 1ex}
 $b_0$ & $-0.821\pm 0.015$ & $-0.822\pm 0.015$\\
 $b_1$ & $~~1.374\pm 0.058$ & $~~1.377\pm 0.057$\\
 $b_2$ & $-0.513\pm 0.086$ & $-0.516\pm 0.083$\\
 $b_3$ & $-1.082\pm 0.056$ & $-1.085\pm 0.059$\\
\hline
 $\chi^2/\mathrm{d.o.f}$ & $76.9/98$ & $76.5/107$\\
\hline\hline
\end{tabular*}
\label{tab:GMp12Fit}
%\vspace{-1ex}
\end{table}

\begin{figure}[t]
    \includegraphics[width=\columnwidth]{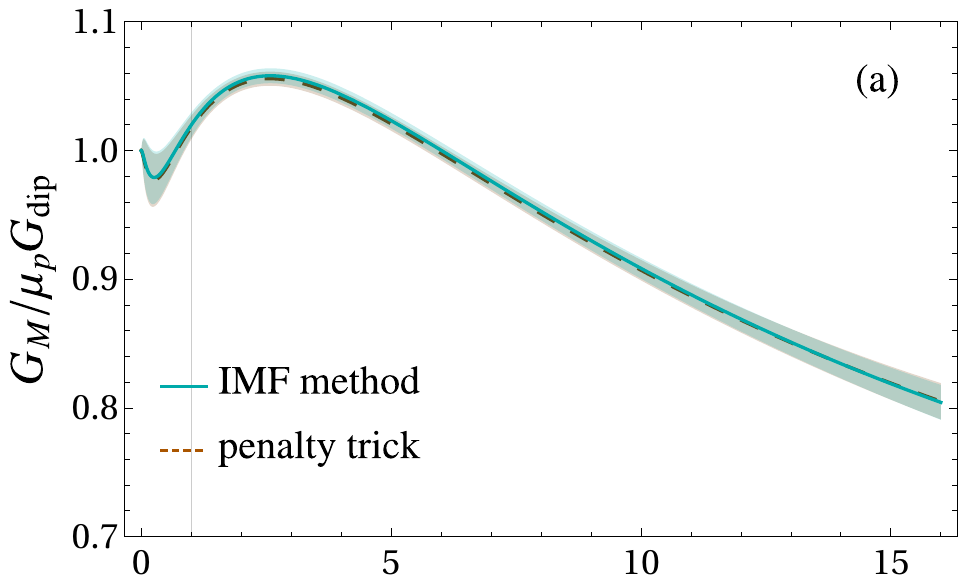}
    \includegraphics[width=\columnwidth]{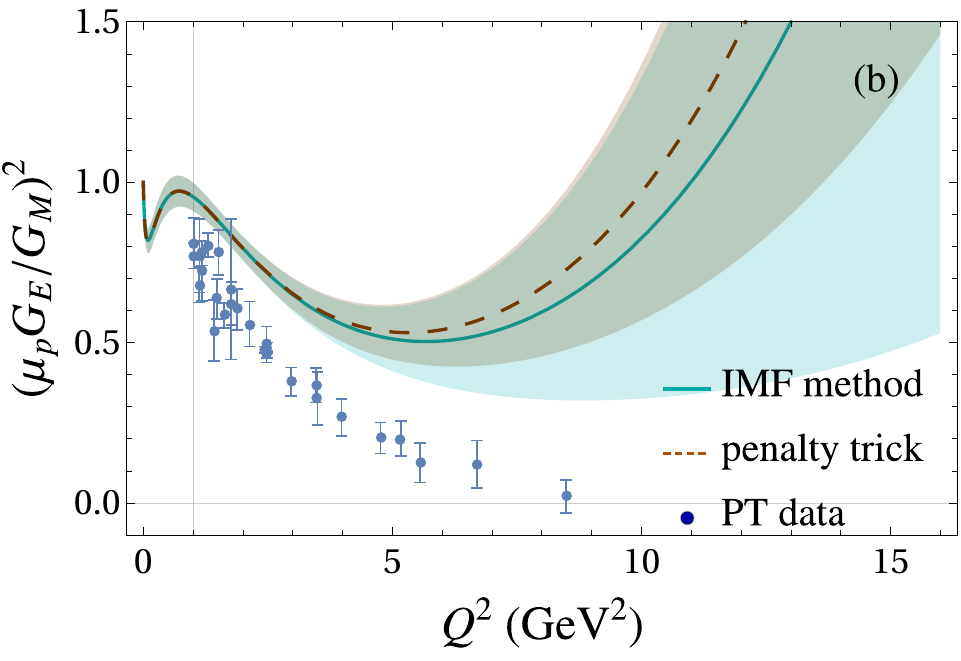}
    \caption{
    The $z$-expansion model fits to the global cross section data from the nine data sets of Ref.~\cite{Christy:2022} with $Q^2<16~\textrm{GeV}^2$.}
    \label{fig:NonlinearFit}
%\vspace{-3ex}
\end{figure}

Using the IMF method with $1500$ replicas per iteration we find convergence within $5$ iterations.
In each iteration about 10\% of replica form factor fits were discarded as they rapidly changed sign in the region $Q^2<1~\textrm{GeV}^2$.
This happens because the model is insensitive to the sign of the form factors, and we have not included any data in this region in the fit.
The $z$-expansion model form factors for both the IMF and penalty trick methods are given in Table~\ref{tab:GMp12Fit} and plotted in Fig.~\ref{fig:NonlinearFit}. 
These fits demonstrate that unless further PT data reveals an upturn, TPE corrections become increasingly important at large $Q^2$.
The relative sizes of the error bands are not surprising.
The Hessian method for approximating parameter errors in nonlinear model fitting is known to be less accurate than bootstrap methods.
We also note that the central value of the penalty trick fit pulls further away from the PT data than the IMF fit does.
Thus it may be the case that the use of the penalty trick method accentuates the disagreement with the PT data.

\section{The importance of correlations}
It is clear that the IMF method achieves fits of the same quality as the penalty trick fits without the need for the additional normalization parameters $n_i$, some of which differ substantially from $1$.
This demonstrates that these additional parameters are simply a means to approximate the correlated nature of the data, and are not physically meaningful on their own.
Correlations enter via the normalization uncertainty due to the fact that if the true value of the normalization is varied the whole data set should vary in tandem. 

This covariance of whole data sets requires extra attention when fitting said data, particularly when faced with the realities of fitting to data subsets.
When fitting with only point-to-point errors, the addition or removal of a data point from the fit will only affect the fit locally near that point.
However, when covariance is of the order of the point-to-point error, truncation and binning of the data set can significantly alter the best fit as the presence of covariance affects the fit function globally. 
In principle, the choice of inclusion or exclusion of a single data point can significantly tense or relax a fit.
It is for this reason we do not perform LT separations invoking the IMF method, as separate linear fits to the relevant data subsets would be blind to the covariance of data at different $Q^2$ values. 

With the same goal in mind, there is often a temptation to treat the normalization factors from the penalty trick as a pseudo model-independent means to rescale data sets for subsequent analysis.
This was done in Ref.~\cite{Christy:2022} to facilitate an LT separation of the rescaled data, giving `model-independent' form factors that could be compared to those obtained directly from the penalty trick fit.
This approach ignores the other correlations between data, and therefore, in our view, does not achieve the goal of improving our knowledge of the true values of the form factors. 
Fitted normalizations are designed to pull the cross section data towards the model, and thus it is no surprise that an LT separation performed on rescaled data gives form factors that are very similar to the those obtained from the penalty trick fit. 

All this furthers a point made by Bernauer~{\em et al.}~\cite{Bernauer:2014} that LT separations are not particularly helpful for fitting when compared to using cross section data.
A global fit appears the only reasonable tool for form factor extraction when normalization uncertainty is present because the total covariance matrix can be incorporated during the fitting procedure.
By contrast, an LT separation is forced to focus on only a subset of the data, ignoring said correlations. 

\section{Summary and Conclusions}
In this work an unbiased global analysis of the proton's elastic form factors was presented.
To perform these fits we introduced the IMF method, an extension of the unbiased $t_0$ fitting method used by the NNPDF collaboration~\cite{Ball:2010}.
We found that unbiased fitting leads to minor improvements to fits of form factor models to the world cross section data without the need for additional normalization parameters for each data set.

The penalty trick, despite its flaws, seems to consistently provide reasonable fits which agree with other experiments.
This begs the question why put in all the extra effort of the IMF method for presumably a small gain?
D’Agostini~\cite{DAgostini:1994} gives a real-world example of how the naive treatment of normalization uncertainties via the penalty trick led to a ‘repulsion’ of the fit from the data, while the chi-square per degree of freedom remained approximately constant.
At the time this was remedied by updating the penalty trick to scale the uncertainties as well as the data, leading to Eq.~(\ref{eq:chi_penalty_updated}).
However, Ball {\em et al.}~\cite{Ball:2010} show this version still fails to be free of bias when combining multiple experimental data sets.
This means there is nothing stopping analogous misbehaviour of fits with the updated penalty trick.

Both the low $Q^2$ global analysis by Bernauer~{\em et al.}~\cite{Bernauer:2014} and recent work on the proton radius~\cite{Lin:2022} use the penalty trick in order to fit their global data sets.
Encouraged by the success of the IMF method, we plan to reanalyze these fits to see if the unbiased fitting leads to any improvements in the extracted values and fits.
Additionally, we plan to perform a combined global analysis of all readily available cross section and PT data, including TPE corrections, in a future work.

\acknowledgements
We thank Alexandr Gramolin for helpful correspondence. This work was supported by the Natural Sciences and Engineering Research Council of Canada.
\vspace{-1ex}
%\bibliography{Refs}
\input{Global.bbl}

\end{document}

%% file: Global.bbl
%Control: production of eprint (-1) disabled
%

%% file: Global.bbl
\begin{thebibliography}{12}%
\makeatletter
\providecommand \@ifxundefined [1]{%
 \@ifx{#1\undefined}
}%
\providecommand \@ifnum [1]{%
 \ifnum #1\expandafter \@firstoftwo
 \else \expandafter \@secondoftwo
 \fi
}%
\providecommand \@ifx [1]{%
 \ifx #1\expandafter \@firstoftwo
 \else \expandafter \@secondoftwo
 \fi
}%
\providecommand \natexlab [1]{#1}%
\providecommand \enquote  [1]{``#1''}%
\providecommand \bibnamefont  [1]{#1}%
\providecommand \bibfnamefont [1]{#1}%
\providecommand \citenamefont [1]{#1}%
\providecommand \href@noop [0]{\@secondoftwo}%
\providecommand \href [0]{\begingroup \@sanitize@url \@href}%
\providecommand \@href[1]{\@@startlink{#1}\@@href}%
\providecommand \@@href[1]{\endgroup#1\@@endlink}%
\providecommand \@sanitize@url [0]{\catcode `\\12\catcode `\$12\catcode
  `\&12\catcode `\#12\catcode `\^12\catcode `\_12\catcode `\%12\relax}%
\providecommand \@@startlink[1]{}%
\providecommand \@@endlink[0]{}%
\providecommand \url  [0]{\begingroup\@sanitize@url \@url }%
\providecommand \@url [1]{\endgroup\@href {#1}{\urlprefix }}%
\providecommand \urlprefix  [0]{URL }%
\providecommand \Eprint [0]{\href }%
\providecommand \doibase [0]{http://dx.doi.org/}%
\providecommand \selectlanguage [0]{\@gobble}%
\providecommand \bibinfo  [0]{\@secondoftwo}%
\providecommand \bibfield  [0]{\@secondoftwo}%
\providecommand \translation [1]{[#1]}%
\providecommand \BibitemOpen [0]{}%
\providecommand \bibitemStop [0]{}%
\providecommand \bibitemNoStop [0]{.\EOS\space}%
\providecommand \EOS [0]{\spacefactor3000\relax}%
\providecommand \BibitemShut  [1]{\csname bibitem#1\endcsname}%
\let\auto@bib@innerbib\@empty
%</preamble>
\bibitem [{\citenamefont {Rosenbluth}(1950)}]{Rosenbluth:1950}%
  \BibitemOpen
  \bibfield  {author} {\bibinfo {author} {\bibfnamefont {M.~N.}\ \bibnamefont
  {Rosenbluth}},\ }\href {\doibase 10.1103/PhysRev.79.615} {\bibfield
  {journal} {\bibinfo  {journal} {Phys. Rev.}\ }\textbf {\bibinfo {volume}
  {79}},\ \bibinfo {pages} {615} (\bibinfo {year} {1950})}\BibitemShut
  {NoStop}%
\bibitem [{\citenamefont {Blunden}\ \emph {et~al.}(2003)\citenamefont
  {Blunden}, \citenamefont {Melnitchouk},\ and\ \citenamefont
  {Tjon}}]{Blunden:2003}%
  \BibitemOpen
  \bibfield  {author} {\bibinfo {author} {\bibfnamefont {P.~G.}\ \bibnamefont
  {Blunden}}, \bibinfo {author} {\bibfnamefont {W.}~\bibnamefont
  {Melnitchouk}}, \ and\ \bibinfo {author} {\bibfnamefont {J.~A.}\ \bibnamefont
  {Tjon}},\ }\href {\doibase 10.1103/PhysRevLett.91.142304} {\bibfield
  {journal} {\bibinfo  {journal} {Phys. Rev. Lett.}\ }\textbf {\bibinfo
  {volume} {91}},\ \bibinfo {pages} {142304} (\bibinfo {year}
  {2003})}\BibitemShut {NoStop}%
\bibitem [{\citenamefont {Guichon}\ and\ \citenamefont
  {Vanderhaeghen}(2003)}]{Guichon:2003}%
  \BibitemOpen
  \bibfield  {author} {\bibinfo {author} {\bibfnamefont {P.~A.~M.}\
  \bibnamefont {Guichon}}\ and\ \bibinfo {author} {\bibfnamefont
  {M.}~\bibnamefont {Vanderhaeghen}},\ }\href {\doibase
  10.1103/PhysRevLett.91.142303} {\bibfield  {journal} {\bibinfo  {journal}
  {Phys. Rev. Lett.}\ }\textbf {\bibinfo {volume} {91}},\ \bibinfo {pages}
  {142303} (\bibinfo {year} {2003})}\BibitemShut {NoStop}%
\bibitem [{\citenamefont {Bernauer}\ \emph {et~al.}(2014)\citenamefont
  {Bernauer} \emph {et~al.}}]{Bernauer:2014}%
  \BibitemOpen
  \bibfield  {author} {\bibinfo {author} {\bibfnamefont {J.~C.}\ \bibnamefont
  {Bernauer}} \emph {et~al.} (\bibinfo {collaboration} {A1}),\ }\href {\doibase
  10.1103/PhysRevC.90.015206} {\bibfield  {journal} {\bibinfo  {journal} {Phys.
  Rev. C}\ }\textbf {\bibinfo {volume} {90}},\ \bibinfo {pages} {015206}
  (\bibinfo {year} {2014})}\BibitemShut {NoStop}%
\bibitem [{\citenamefont {Ball}\ \emph {et~al.}(2010)\citenamefont {Ball},
  \citenamefont {Del~Debbio}, \citenamefont {Forte}, \citenamefont {Guffanti},
  \citenamefont {Latorre}, \citenamefont {Rojo},\ and\ \citenamefont
  {Ubiali}}]{Ball:2010}%
  \BibitemOpen
  \bibfield  {author} {\bibinfo {author} {\bibfnamefont {R.~D.}\ \bibnamefont
  {Ball}}, \bibinfo {author} {\bibfnamefont {L.}~\bibnamefont {Del~Debbio}},
  \bibinfo {author} {\bibfnamefont {S.}~\bibnamefont {Forte}}, \bibinfo
  {author} {\bibfnamefont {A.}~\bibnamefont {Guffanti}}, \bibinfo {author}
  {\bibfnamefont {J.~I.}\ \bibnamefont {Latorre}}, \bibinfo {author}
  {\bibfnamefont {J.}~\bibnamefont {Rojo}}, \ and\ \bibinfo {author}
  {\bibfnamefont {M.}~\bibnamefont {Ubiali}} (\bibinfo {collaboration}
  {NNPDF}),\ }\href {\doibase 10.1007/JHEP05(2010)075} {\bibfield  {journal}
  {\bibinfo  {journal} {JHEP}\ }\textbf {\bibinfo {volume} {05}},\ \bibinfo
  {pages} {075} (\bibinfo {year} {2010})}\BibitemShut {NoStop}%
\bibitem [{\citenamefont {Gramolin}\ and\ \citenamefont
  {Nikolenko}(2016)}]{Gramolin:2016}%
  \BibitemOpen
  \bibfield  {author} {\bibinfo {author} {\bibfnamefont {A.~V.}\ \bibnamefont
  {Gramolin}}\ and\ \bibinfo {author} {\bibfnamefont {D.~M.}\ \bibnamefont
  {Nikolenko}},\ }\href {\doibase 10.1103/PhysRevC.93.055201} {\bibfield
  {journal} {\bibinfo  {journal} {Phys. Rev. C}\ }\textbf {\bibinfo {volume}
  {93}},\ \bibinfo {pages} {055201} (\bibinfo {year} {2016})}\BibitemShut
  {NoStop}%
\bibitem [{\citenamefont {Walker}\ \emph {et~al.}(1994)\citenamefont {Walker}
  \emph {et~al.}}]{Walker:1994}%
  \BibitemOpen
  \bibfield  {author} {\bibinfo {author} {\bibfnamefont {R.~C.}\ \bibnamefont
  {Walker}} \emph {et~al.},\ }\href {\doibase 10.1103/PhysRevD.49.5671}
  {\bibfield  {journal} {\bibinfo  {journal} {Phys. Rev. D}\ }\textbf {\bibinfo
  {volume} {49}},\ \bibinfo {pages} {5671} (\bibinfo {year}
  {1994})}\BibitemShut {NoStop}%
\bibitem [{\citenamefont {Andivahis}\ \emph {et~al.}(1994)\citenamefont
  {Andivahis} \emph {et~al.}}]{Andivahis:1994}%
  \BibitemOpen
  \bibfield  {author} {\bibinfo {author} {\bibfnamefont {L.}~\bibnamefont
  {Andivahis}} \emph {et~al.},\ }\href {\doibase 10.1103/PhysRevD.50.5491}
  {\bibfield  {journal} {\bibinfo  {journal} {Phys. Rev. D}\ }\textbf {\bibinfo
  {volume} {50}},\ \bibinfo {pages} {5491} (\bibinfo {year}
  {1994})}\BibitemShut {NoStop}%
\bibitem [{\citenamefont {Christy}\ \emph {et~al.}(2022)\citenamefont {Christy}
  \emph {et~al.}}]{Christy:2022}%
  \BibitemOpen
  \bibfield  {author} {\bibinfo {author} {\bibfnamefont {M.~E.}\ \bibnamefont
  {Christy}} \emph {et~al.},\ }\href {\doibase 10.1103/PhysRevLett.128.102002}
  {\bibfield  {journal} {\bibinfo  {journal} {Phys. Rev. Lett.}\ }\textbf
  {\bibinfo {volume} {128}},\ \bibinfo {pages} {102002} (\bibinfo {year}
  {2022})}\BibitemShut {NoStop}%
\bibitem [{\citenamefont {Lee}\ \emph {et~al.}(2015)\citenamefont {Lee},
  \citenamefont {Arrington},\ and\ \citenamefont {Hill}}]{Lee:2015}%
  \BibitemOpen
  \bibfield  {author} {\bibinfo {author} {\bibfnamefont {G.}~\bibnamefont
  {Lee}}, \bibinfo {author} {\bibfnamefont {J.~R.}\ \bibnamefont {Arrington}},
  \ and\ \bibinfo {author} {\bibfnamefont {R.~J.}\ \bibnamefont {Hill}},\
  }\href {\doibase 10.1103/PhysRevD.92.013013} {\bibfield  {journal} {\bibinfo
  {journal} {Phys. Rev. D}\ }\textbf {\bibinfo {volume} {92}},\ \bibinfo
  {pages} {013013} (\bibinfo {year} {2015})}\BibitemShut {NoStop}%
\bibitem [{\citenamefont {D'Agostini}(1994)}]{DAgostini:1994}%
  \BibitemOpen
  \bibfield  {author} {\bibinfo {author} {\bibfnamefont {G.}~\bibnamefont
  {D'Agostini}},\ }\href {\doibase 10.1016/0168-9002(94)90719-6} {\bibfield
  {journal} {\bibinfo  {journal} {Nucl. Instrum. Meth. A}\ }\textbf {\bibinfo
  {volume} {346}},\ \bibinfo {pages} {306} (\bibinfo {year}
  {1994})}\BibitemShut {NoStop}%
\bibitem [{\citenamefont {Lin}\ \emph {et~al.}(2022)\citenamefont {Lin},
  \citenamefont {Hammer},\ and\ \citenamefont {Mei\ss{}ner}}]{Lin:2022}%
  \BibitemOpen
  \bibfield  {author} {\bibinfo {author} {\bibfnamefont {Y.-H.}\ \bibnamefont
  {Lin}}, \bibinfo {author} {\bibfnamefont {H.-W.}\ \bibnamefont {Hammer}}, \
  and\ \bibinfo {author} {\bibfnamefont {U.-G.}\ \bibnamefont {Mei\ss{}ner}},\
  }\href {\doibase 10.1103/PhysRevLett.128.052002} {\bibfield  {journal}
  {\bibinfo  {journal} {Phys. Rev. Lett.}\ }\textbf {\bibinfo {volume} {128}},\
  \bibinfo {pages} {052002} (\bibinfo {year} {2022})}\BibitemShut {NoStop}%
\end{thebibliography}
